\documentclass[aps,twocolumn,groupeaddress]{revtex4}

\newcommand{\dif}{\mathrm{d}}

\usepackage{graphicx}

\usepackage{amsmath}

\usepackage{amsfonts}

\usepackage{dcolumn}% Align table columns on decimal point

\usepackage{bm}% bold math

\begin{document}

\preprint{LA-UR-06-5877}

\title{Gaussian Models for the Statistical Thermodynamics of Liquid
Water}% Force line breaks with \\

\author{J. K. Shah}\altaffiliation{Department of Chemical and Biomolecular
Engineering, Ohio State University, Columbus OH 43210 USA}

\author{D. Asthagiri}
\altaffiliation{Department of Chemical and Biomolecular Engineering,
Johns Hopkins University, Baltimore MD 21218 USA}

\author{L. R. Pratt}
\altaffiliation{Theoretical Division, Los Alamos National Laboratory,
Los Alamos NM 87545 USA}

\author{M. E. Paulaitis$^\ast$}

\date{\today}
             
\begin{abstract} A gaussian distribution of binding energies, but
conditioned to exploit generally available information on packing in
liquids, provides a statistical-thermodynamic theory of liquid water
that is structurally non-committal, molecularly realistic, and
surprisingly accurate. Neglect of fluctuation contributions to this
gaussian model yields a mean-field theory that produces useless results.
A refinement that accounts for sharper-than-gaussian behavior at high
binding energies recognizes  contributions from a discrete number of
water molecules and permits a natural matching of numerically exact
results. These gaussian models, which can be understood as vigorous
simplifications  of quasi-chemical theories, are applicable to aqueous
environments where the utility of structural models based on geometrical
considerations of water hydrogen bonding have not been established. 
\end{abstract}

\maketitle

%\section*{Introduction} 
% We show that a realistic statistical thermodynamic theory of liquid water can be based upon a conditional gaussian distribution of binding energies that is structurally non-committal, where the condition exploits generally available information on packing in liquids.   These gaussian models can are vigorous simplifications  of quasi-chemical theories \cite{Asthagiri:2003,PaliwalA.:Anamp}, and are applicable to aqueous environments where the utility of a structural picture  of liquid water has not been established.  Examples are found in the aqueous materials that may be encountered in planetary exploration guided by the \emph{follow the water} strategy \cite{HubbardGS:Folwnp}, including proposed alternatives to water as a matrix for life \cite{Varenna}; further examples include confined water encountered in biophysical systems, such as water in protein cavities, \cite{CollinsMD:Coowfn} pores, \cite{pande:jacs06}, or interfaces \cite{paliwal:bj05}. This model should also provide helpful characterization of simulated metastable solutions \cite{PaschekD:Howlta}.

\begin{figure}
	\begin{center}
    \includegraphics[width=3.0in]{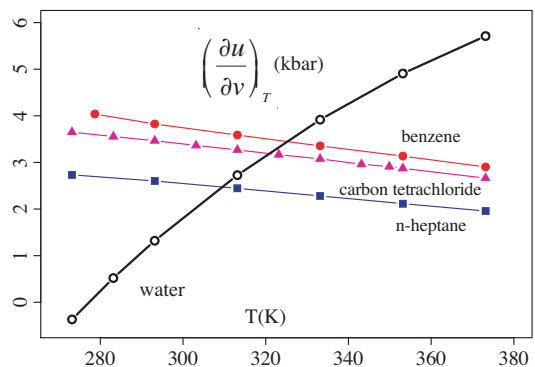}
    \caption{$\left(\partial u/\partial v\right)_T$ for several solvents
    as a function of temperature along the vapor saturation curve.  For
    van der Waals liquids $\left(\partial u/\partial v\right)_T \approx
    a \rho^2$.  Organic solvents conform to this expectation, but water 
    is qualitatively different.\label{fig:internal_pressure}  The data
    are from \cite{Rowlinson}.}  
	\end{center}
\end{figure}

A widely accepted molecular-scale understanding of liquid water under
physiological conditions has evolved over recent decades based upon the
concept of promiscuous hydrogen-bonding that results in a thoroughly
networked fluid \cite{STILLINGERFH:Watr}. This view rests upon extensive
molecular-scale simulation validated with traditional experimentation,
and communicated with molecular-graphics tools. But water is a peculiar
liquid.  For example,  the van~der~Waals theory \cite{Widom:67}, which
provides the firmest basis for theories of simple liquids, is
unsatisfactory for liquid water. Fig.~\ref{fig:internal_pressure} shows
one experimental demonstration of that point.  The foremost task in
applying molecular statistical mechanical theory to liquid water is to
address the equation-of-state distinctions exemplified in
Fig.~\ref{fig:internal_pressure} on the basis of realistic
intermolecular interactions. One theoretical approach is to accept the
voluminous data that can be generated in a typical, realistic molecular
simulation, but to craft a concise, quantitative statistical description
of the basic thermodynamic characteristics
\cite{HummerG:Anitm,Ashbaugh:2006}.  As exemplified below, those
statistical theories can be concise  indeed, and general in scope.

The focus here is analyzing the probability density distribution,
$p(\varepsilon)$, of binding energies, $\varepsilon$, exhibited by a
water molecule in  liquid  water. Thermodynamic properties are typically
sensitive to the  tails of this distribution,   As a topical example,
note that the population of weakly bound water molecules in liquid water
can be decisive in filling transitions of carbon nanotubes
\cite{HummerG:Watcth}.   Thus, it can be helpful to have a clear idea
how those weakly bound populations can be analyzed, and a focus of this
work is the analysis of $p(\varepsilon)$.

%\section*{Statistical Thermodynamic Theory}

Our motivation for the present analysis is the observation of severely
non-gaussian $p(\varepsilon)$ in cases where repulsive interactions are
prominent contributors \cite{CF4}.  In such cases, conditioning to
separate out effects of repulsive interactions was found to yield
conditional distributions, $p\left(\varepsilon \vert r \ge
\lambda\right)$, that were accurately gaussian. The idea is to account
separately for close molecular encounters; then the direct statistical
problem of evaluating the distribution of binding energies need only
consider  the fraction of the sample for which the distance to the
nearest solvent molecule center, $r$, is greater than the conditioning
radius $\lambda$.  That fraction is $p\left(r \ge \lambda\right)$, the
marginal probability.  In previous quasi-chemical treatments, the
marginal probability $p\left(r \ge \lambda\right)$ was denoted by $x_0$
\cite{Asthagiri:2003,PaliwalA.:Anamp,CF4,Paulaitis:2002}.

\begin{figure}[t]
%\begin{center}
\includegraphics[width=3.5in]{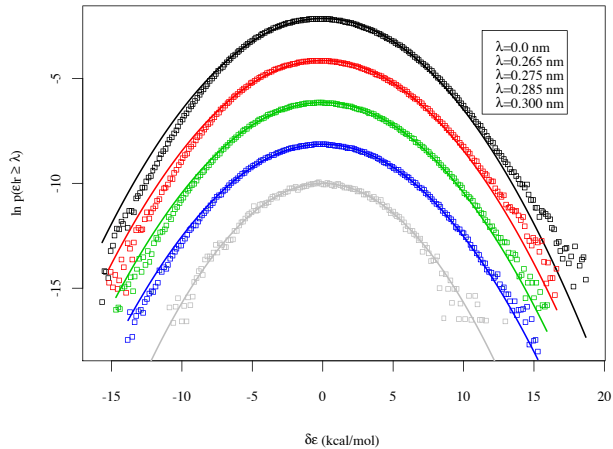}
%\end{center}
\caption{\label{Fig:comparison_rcut_full_T300} Probability density 
$p(\varepsilon \vert r \ge \lambda)$ of the binding energy of a water
molecule in liquid water at 298 K.  $\lambda$ =  0.0, \ldots 0.300~nm,
from  top to bottom with successive results shifted
incrementally shifted downward by 2 for clarity.  The solid lines are
the gaussian model for each data set.}
\end{figure}

To follow that path, we seek $\mu^\mathrm{{ex}}$, the chemical potential
of water in excess of the ideal contribution at the same density and
temperature.  On the basis of simulation data, we consider evaluating
$\mu^\mathrm{{ex}}_\mathrm{{HS}}$, the excess chemical potential of a
hard sphere solute, relative to $\mu^\mathrm{{ex}}$.  The potential
distribution theorem (PDT) \cite{CF4,Beck:2006,Paulaitis:2002} then
yields
\begin{eqnarray}
e^{-\beta(\mu^\mathrm{{ex}}_\mathrm{{HS}} - \mu^\mathrm{{ex}})} = 
p\left(r \ge \lambda\right) 
\int\limits_{-\infty}^{+\infty} p\left(\varepsilon \vert r \ge \lambda\right) e^{\beta\varepsilon} \dif\varepsilon ~.
\label{eq:derivation}
\end{eqnarray}
The thermodynamic temperature is $T=1/k_B\beta$ where $k_B$ is the
Boltzmann's constant.  Since $\mu^\mathrm{{ex}}_\mathrm{{HS}}$ is known
\cite{Ashbaugh:2006}, Eq.~\eqref{eq:derivation} gives
$\mu^\mathrm{{ex}}$.  We regard this conditioning as a {\it
regularization} of the statistical problem embodied in
Eq.~\eqref{eq:derivation} when $\lambda\rightarrow 0$, which is
practically impossible on the basis of a direct, single calculation. 
After regularization, the statistical problem becomes merely difficult.
A gaussian distribution model for $p\left(\varepsilon\vert r \ge \lambda
\right)$ should be accurate when $\lambda\rightarrow\infty$, since then
many solution elements will make small, weakly-correlated contributions
to $\varepsilon$.  The marginal probability $p\left(r \ge
\lambda\right)$ becomes increasingly difficult to evaluate as $\lambda$
becomes large, however. For $\lambda$ on the order of molecular length
scales typical of dense liquids, a simple gaussian model would accept
some approximation error as the price for manageable statistical error.
If $p\left(\varepsilon\vert r \ge \lambda \right)$ is modeled by a
gaussian of mean $\left<\varepsilon\vert r \ge \lambda \right>$ and
variance $\left < \delta \varepsilon ^2 \vert r \ge \lambda \right >$,
then
\begin{multline}
\mu^\mathrm{{ex}} - \mu^\mathrm{{ex}}_\mathrm{{HS}} 
- kT \ln p\left(r \ge \lambda\right) 
-  \left<\varepsilon\vert r \ge \lambda \right>  = \\
\frac{1}{2 kT}\left < \delta \varepsilon ^2 \vert r \ge \lambda \right >
~.
\label{eq:working_eq_gaussian}
\end{multline}
%%%%%%%%%%%%%%%%%%%%%%%%%%%%%%%%%%%%%%%%%%%%%%%%%%%%%%%%%%
This simple  model motivates the following analyses.

\begin{figure}[b]
\begin{center}
\includegraphics[width=3.5in]{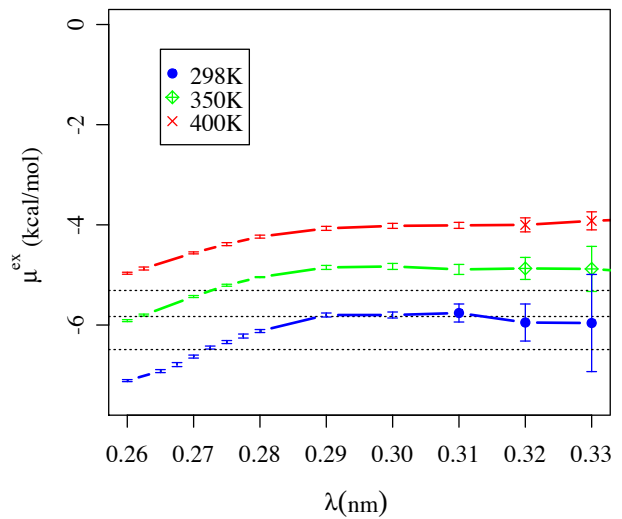}
\end{center}
\caption{Dependence of the free energy $\mu^{\mathrm{ex}}$ predicted by
the gaussian model on the conditioning radius $\lambda$. The horizontal
dotted lines are the numerically exact results. The error bars indicate
approximate 95\% confidence intervals.  \label{fig:murdep}}
\end{figure}

%\section*{Simulation Data}
To test these ideas, simulation data for liquid water was generated at 298, 350, and 400 K and 1 bar using methods described in \cite{PaliwalA.:Anamp}.  The hard-sphere excess chemical potential was obtained from
\cite{Ashbaugh:2006}.  The distributions observed for $T$ = 298 K are shown in
Fig~\ref{Fig:comparison_rcut_full_T300}. Table~\ref{tab:results} collects the individual terms for the gaussian model, Eq.~\eqref{eq:working_eq_gaussian},  at each temperature. The observed dependence on $\lambda$ of the free energy at each temperature is shown in   Fig~\ref{fig:murdep}.

%\section*{Discussion}

Fig.~\ref{Fig:comparison_rcut_full_T300} shows that the unconditioned
distribution $p\left(\varepsilon\right)$ displays positive skew, but the
conditioning diminishes that skew perceptibly, as expected.
$p\left(\varepsilon\vert r \ge \lambda \right)$ is least skewed for the
largest $\lambda$, though the sample used is smaller by the fraction
$p\left(r \ge \lambda\right) $, and thus less of the tail region is available
for examination as $\lambda$ becomes larger.

\begin{table*}
\begin{center}
\caption{Free energy contributions in kcal/mol associated with the
gaussian model.  The bottom value of the right-most column  at each
temperature gives the corresponding free energy evaluated by the
histogram overlap method.  \label{tab:results}}

\begin{tabular}{|c|l|ccrcc|c|}
\hline
	   T(K)  
	&  $\lambda$ (nm)  
	&  $\mu^{\mathrm{ex}}_{\mathrm{HS}}\left(\lambda \right)$ 
	&  $+ kT\ln p\left(r\ge \lambda  \right)$
	&  $+ \left\langle \varepsilon\vert r\ge\lambda \right\rangle$ 
	&  $+\left\langle \delta \varepsilon^2\vert r\ge\lambda \right\rangle/2kT$  
	&  $= \mu^{\mathrm{ex}}$   \\ \hline
	   298
	& 0.2600 & 2.80 & $-0.04$ & $-19.74$  & +9.87 & = $-7.11 \pm$ 0.02   \\
	& 0.2650 & 2.99 & $-0.13$ & $-19.68$  & +9.93 & = $-6.89 \pm$ 0.03  \\
    & 0.2675 & 3.09 & $-0.20$ & $-19.59$  & +9.97 & = $-6.73 \pm$ 0.04   \\
    & 0.2700 & 3.19 & $-0.31$ & $-19.46$  & +9.98 & = $-6.60 \pm$ 0.03  \\
 	& 0.2725 & 3.29 & $-0.44$ & $-19.27$  & +9.97 & = $-6.45 \pm$ 0.03  \\ 
    & 0.2750 & 3.40 & $-0.60$ & $-19.03$  & +9.92 & = $-6.31 \pm$ 0.03  \\
    & 0.2775 & 3.50 & $-0.78$ & $-18.73$  & +9.83 & = $-6.18 \pm$ 0.04   \\
    & 0.2800 & 3.61 & $-0.98$ & $-18.39$  & +9.71 & = $-6.05 \pm$ 0.03  \\
    & 0.2900 & 4.07 & $-1.96$ & $-16.75$  & +8.89 & = $-5.75 \pm$ 0.04   \\
    & 0.3000 & 4.56 & $-3.09$ & $-14.93$  & +7.77 & = $-5.69 \pm$ 0.06  \\
    & 0.3100 & 5.05 & $-4.27$ & $-13.21$  & +6.67 & = $-5.67 \pm$ 0.18  \\
    & 0.3200 & 5.61 & $-5.45$ & $-11.65$  & +5.54 & = $-5.95 \pm$ 0.37  \\
    & 0.3300 & 6.20 & $-6.64$ & $-10.30$  & +4.78 & = $-5.96 \pm$ 0.97  \\
    &        &      &         &         &       &   $-6.49$    \\ \hline
       350  
	& 0.2600 & 3.12 & $-0.05$ & $-18.43$ & +9.44 & = $-5.92 \pm$ 0.02  \\
    & 0.2625 & 3.23 & $-0.09$ & $-18.41$ & +9.46 & = $-5.81 \pm$ 0.02  \\
    & 0.2700 & 3.55 & $-0.33$ & $-18.14$ & +9.47 & = $-5.45 \pm$ 0.02  \\
    & 0.2750 & 3.77 & $-0.62$ & $-17.73$ & +9.35 & = $-5.23 \pm$ 0.02  \\
    & 0.2800 & 4.00 & $-0.99$ & $-17.15$ & +9.10 & = $-5.04 \pm$ 0.01  \\
    & 0.2900 & 4.50 & $-1.92$ & $-15.67$ & +8.24 & = $-4.85 \pm$ 0.04  \\
    & 0.3000 & 5.02 & $-3.00$ & $-14.05$ & +7.20 & = $-4.83 \pm$ 0.06  \\
    & 0.3100 & 5.58 & $-4.13$ & $-12.48$ & +6.14 & = $-4.89 \pm$ 0.10  \\
    & 0.3200 & 6.18 & $-5.28$ & $-11.01$ & +5.24 & = $-4.87 \pm$ 0.22  \\
    & 0.3300 & 6.80 & $-6.41$ & $-9.74$  & +4.47 & = $-4.88 \pm$ 0.45  \\
%    & 0.3500 & 8.12 & $-8.61$ & $-7.39$  & +2.82 & = $-5.06 \pm$ 1.74  \\ 
    &        &      &         &        &         & $-5.83$   \\ \hline
       400  
	& 0.2600 & 3.30 & $-0.06$ & $-17.19$ & +8.97 & = $-4.98 \pm$ 0.02  \\
    & 0.2625 & 3.40 & $-0.10$ & $-17.16$ & +8.99 & = $-4.87 \pm$ 0.03  \\
    & 0.2700 & 3.74 & $-0.35$ & $-16.89$ & +8.95 & = $-4.55 \pm$ 0.02  \\
    & 0.2750 & 3.96 & $-0.63$ & $-16.49$ & +8.80 & = $-4.36 \pm$ 0.03   \\
    & 0.2800 & 4.20 & $-0.98$ & $-15.96$ & +8.52 & = $-4.22 \pm$ 0.03  \\
    & 0.2900 & 4.71 & $-1.87$ & $-14.60$ & +7.69 & = $-4.07 \pm$ 0.05  \\
    & 0.3000 & 5.25 & $-2.89$ & $-13.10$ & +6.72 & = $-4.02 \pm$ 0.05  \\
    & 0.3100 & 5.82 & $-3.97$ & $-11.64$ & +5.78 & = $-4.01 \pm$ 0.06  \\
    & 0.3200 & 6.42 & $-5.06$ & $-10.31$ & +4.95 & = $-4.00 \pm$ 0.14  \\
    & 0.3300 & 7.05 & $-6.12$ & $-9.12$  & +4.27 & = $-3.92 \pm$ 0.18  \\
%    & 0.3500 & 8.39 & $-8.15$ & $-7.14$ & +3.08 &  = $-3.82 \pm$ 0.47  \\ 
%    & 0.3600 & 9.10 & $-9.12$ & $-6.32$ & +2.81 &  = $-3.53 \pm$ 0.86  \\ 
%    & 0.3700 & 9.83 & $-10.1$ & $-5.74$ & +2.26 &  = $-3.75 \pm$ 2.12  \\ 
    &        &      &         &          &         & $-5.31$   \\ \hline
\end{tabular}
\end{center}
\end{table*}

The conditioning affects both the high-$\varepsilon$ and
low-$\varepsilon$ tails of these distributions.  The mean binding energy
$\left\langle \varepsilon\vert r>\lambda \right\rangle$ increases with
increasing $\lambda$ [Table~\ref{tab:results}], so we conclude that the
conditioning eliminates atypical low-$\varepsilon$, well-bound
configurations more than high-$\varepsilon$ configurations that reflect
less favorable interactions.   Nevertheless, because of the exponential
weighting of the integrand of Eq.~\eqref{eq:derivation} and because the 
variances are large, the high-$\varepsilon$ side of the distributions
is overwhelmingly the more significant in this free energy prediction.

Conversely, the fluctuation contribution exhibits a broad maximum for $\lambda <$ 0.29~nm, after which this contribution decreases steadily with increasing $\lambda$ [Table~\ref{tab:results}].  Evidently water molecules closest to the distinguished
molecule,  \emph{i.e.,} those closer  than the principal maximum of
oxygen-oxygen radial distribution function, don't contribute importantly
to the fluctuations. This is consistent with a quasi-chemical picture in
which a water molecule and its nearest neighbors have a definite
structural integrity.

The magnitude of the individual contributions to $\mu^\mathrm{{ex}}$ are of the same
order as the net free energy; the mean binding energies are
larger than that, as are the  variance contributions in some cases.  The
variance contributions are about half as large as the mean binding
energies, with opposite sign. It is remarkable and significant,
therefore, that the net free energies at 298 K are within roughly 12\%
of the numerically exact value computed by the histogram-overlap method.
%for all the $\lambda$ values considered. 
The discrepancies at the higher
temperatures are larger, and we  will return to  that point. A
mean-field-like approximation that neglects fluctuations produces
useless results. 

We note that $p(r \ge \lambda)\approx$ 1 for the smallest values of $\lambda$ in Table~\ref{tab:results}. % We note that naive expressions of the {\it inverse} form of the PDT are ill-defined for hard-core interactions specifically \cite{HendersonJR:Invpdt}, though the quasi-chemical approaches found one natural circumvention of that problem \cite{PrattLR:Quatst,PrattLawrenceR.:Selmft}.  
This leads to the awkward point that if $kT\ln{p(r \ge \lambda)}$ is zero, then the hard-sphere contribution $\mu^\mathrm{{ex}}_\mathrm{{HS}}$ is ill-defined.  As a general matter, the sum $\mu^\mathrm{{ex}}_\mathrm{{HS}}+kT\ln{p(r \ge \lambda)}$ cannot be identified as a hard-sphere contribution.  Since these terms have opposite signs, the net value can be zero or negative, and those possibilities are realized [Table~\ref{tab:results}].  To define the hard-sphere contribution more generally, we require $\mu^\mathrm{{ex}}_\mathrm{{HS}}$ to be continuous as $\lambda$ decreases, such that $p(r \ge \lambda)\rightarrow$ 1. All other terms of Eq.~~\eqref{eq:working_eq_gaussian} will be independent of $\lambda$ for values smaller than that, and we will require that of $\mu^\mathrm{{ex}}_\mathrm{{HS}}$ also.   %It is a temperature-dependent curiosity that the hard-sphere contribution and the back-filling contribution $kT\ln{p(r \ge \lambda)}$ balance for $\lambda \gtrsim$ 0.32~nm with larger differences found for  higher temperatures.

From Fig.~\ref{fig:murdep}, we see that $\lambda >$ 0.30~nm clearly identifies a larger-size regime where the variation of the free energy with $\lambda$ is not statistically significant. Although we
anticipate a decay toward the numerically exact value for $\lambda
\rightarrow \infty$, the statistical errors become unmanageable for
values of $\lambda$ much larger than 0.30~nm.   When $\lambda$
= 0.30~nm a significant skew  in $p(\varepsilon \vert r \ge \lambda)$ is
not observed, as already noted with
Fig.~\ref{Fig:comparison_rcut_full_T300}.  The predicted free energy
$\mu^\mathrm{{ex}}$ is then distinctly above the numerically exact
value, suggesting that the gaussian model predicts too much weight in
the high-$\varepsilon$ tail. We hypothesize that this
sharper-than-gaussian tail behavior is due to the fact that a finite number of molecules make discrete contributions to the
net $\varepsilon$ in this tail
region. A model distribution exhibiting this distinction is
\begin{multline}
p\left(\varepsilon \vert r \ge \lambda\right)
 = \\
 \left\{\prod\limits_{k=1}^n \int \pi\left(\varepsilon_k\vert r \ge \lambda\right)\right\}
 \delta\left(\varepsilon - \sum\limits_{k=1}^n\varepsilon_k\right)
 \dif\varepsilon_1 \ldots \dif \varepsilon_n~,
 \label{eq:compose}
\end{multline}
with $\pi\left(\varepsilon_k\vert r \ge \lambda\right)$ an elementary
distribution with a sharp cut-off.  The plug density
\begin{eqnarray}
\pi\left(\varepsilon_k\vert r \ge \lambda\right)
 = \left(\frac{1}{\Delta\varepsilon}\right)
 H\left(\frac{\Delta\varepsilon}{2}
 -
 \left\vert \varepsilon_k - 
 	\frac{\left\langle \varepsilon\vert r\ge\lambda \right\rangle}
 		{n}
 	\right\vert\right),
\label{eq:plug}
\end{eqnarray}
with $H\left(x\right)$ the Heaviside function, is an example.  This function is
non-zero over an $\varepsilon$-width of 
\begin{eqnarray}
\Delta\varepsilon=\sqrt{12
\left < \delta \varepsilon ^2 \vert r \ge \lambda \right > /n}
\label{eq:deps}
\end{eqnarray}
when the parameterization is set so that Eq.~\eqref{eq:compose} is
consistent with the previous notation.  This leaves the number $n$ as a
single further parameter;  a larger $n$
indicates a larger range of gaussian behavior for $p(\varepsilon \vert r \ge \lambda)$.  With the example
Eq.~\eqref{eq:plug} the evaluation of the thermodynamic result
is elementary, and amounts to the replacement
\begin{eqnarray}
\left < \delta \varepsilon ^2 \vert r \ge \lambda \right >
\leftarrow
%nkT\ln \left\lbrack {\frac{{\sinh \left( {\sqrt {3\beta ^2\left < \delta \varepsilon ^2 \vert r \ge \lambda \right > /n} } \right)}}{{\sqrt {3\beta ^2 \left < \delta \varepsilon ^2 \vert r \ge \lambda \right >/n} }}} \right\rbrack
\left < \delta \varepsilon ^2 \vert r \ge \lambda \right >
\left(1 - \frac{\beta^2\left < \delta \varepsilon ^2 \vert r \ge \lambda \right >}{10 n}\right)
\label{eq:discrete}
\end{eqnarray}
in Eq.~\eqref{eq:working_eq_gaussian} for large $n$. % It is essential, however, to recognize that the \emph{sharper-than-gaussian} hypothesis means sharper than the single gaussian of Eq.~\eqref{eq:working_eq_gaussian}.

Since we exploit a single further datum, a single parameter exhausts
that information. When $\lambda$ = 0.30~nm, the values of $n$ that  fit
the numerically exact results are 21, 11, and 6 at $T$= 298, 350, and
400~K, respectively.  These values are reasonable as indications of the
discrete number of proximal water molecules that dominate solvent
interactions with the distinguished molecule.  

It is essential, however, to recognize that the
\emph{sharper-than-gaussian} hypothesis means sharper than the single
gaussian of Eq.~\eqref{eq:working_eq_gaussian}.  A natural path for improvement of these results would be a multi-gaussian model, as was developed for
careful evaluations of the electrostatic contribution to the free energy of
liquid water \cite{Hummer:1995, Hummer:1997}. But some further  points, which touch upon similarities of the multi-gaussian and quasi-chemical approaches, must be kept in mind.  First, we
are concerned here with contributions from a defined \emph{outer-shell}
region, the direct inner-shell contributions have been relegated to the
$ kT \ln p\left(r \ge \lambda\right)$ term of
Eq.~\eqref{eq:working_eq_gaussian}.   Second, the multi-gaussian approach requires choosing a
variable to stratify the distribution;  in the quasi-chemical approach
that variable is the number of occupants of the defined shell \cite{CPMS}, presumably an {\it inner outer-shell} region here.  Since
this refinement has exhausted the data here, we don't
pursue those more refined approaches.   But we also recognize that the initial model, Eq.~\eqref{eq:working_eq_gaussian}, is a compact, simple implementation of quasi-chemical ideas.

The values of $n$ are found to correlate positively with
the variance $\left < \delta \varepsilon ^2 \vert r \ge \lambda \right
>$, such that $\Delta\varepsilon$ [Eq.~\eqref{eq:deps}] is
only weakly dependent on $\lambda$.  At $T$ = 298~K,
$\Delta\varepsilon \approx$ 2~kcal/mol, independent of $\lambda$. 
At the higher temperatures, $\Delta\varepsilon$ is
1-2~kcal/mol larger, and has a noticeable, linear
dependence on $\lambda$.  These empirical energy parameters are of
reasonable magnitude by comparison with hydrogen-bond energies, and they do
not correspond to weak interactions on a thermal scale.

Though these theoretical developments were unanticipated, it is
possible to make some connection to classical theories.
Assume that the
$N$-molecule fluid can be satisfactorily described by a
pair-decomposable potential energy function.   Then a gaussian model for
a joint distribution of binding energies $\varepsilon_1$ and
$\varepsilon_2$, of molecules 1 and 2, respectively, predicts
\begin{eqnarray}
\beta^2\left< \delta \varepsilon_1\delta \varepsilon_2\vert 1,2 \right>
= \ln y\left(1,2\right)~,
\label{eq:RPA}
\end{eqnarray}
where $y\left(1,2\right)$ is the two-molecule indirect distribution
function \cite{Hansen-McDonald}, and the average is \emph{conditional} 
upon location of molecules at (1,2).  A point of general interest is
that  Eq.~\eqref{eq:RPA} is a signature of the random-phase family of
approximations, \emph{e.g.} the Debye-H\"{u}ckel theory. It is also
interesting that Eq.~\eqref{eq:RPA} does not express a conventional
mean-field contribution. However, if the molecules considered are
significantly different, such a relation then is expected to resemble mean-field contributions of the classical type.

This work was carried out under the auspices of the National Nuclear
Security Administration of the U.S. Department of Energy at Los Alamos
National Laboratory under Contract No. DE-AC52-06NA25396. Financial
support from the National Science Foundation (BES0518922) is gratefully
acknowledged.  %LA-UR-06-5877

%\bibliography{references}% Produces the bibliography via BibTeX.

\begin{thebibliography}{15}
\expandafter\ifx\csname natexlab\endcsname\relax\def\natexlab#1{#1}\fi
\expandafter\ifx\csname bibnamefont\endcsname\relax
  \def\bibnamefont#1{#1}\fi
\expandafter\ifx\csname bibfnamefont\endcsname\relax
  \def\bibfnamefont#1{#1}\fi
\expandafter\ifx\csname citenamefont\endcsname\relax
  \def\citenamefont#1{#1}\fi
\expandafter\ifx\csname url\endcsname\relax
  \def\url#1{\texttt{#1}}\fi
\expandafter\ifx\csname urlprefix\endcsname\relax\def\urlprefix{URL }\fi
\providecommand{\bibinfo}[2]{#2}
\providecommand{\eprint}[2][]{\url{#2}}

\bibitem[{\citenamefont{Stillinger}(1980)}]{STILLINGERFH:Watr}
\bibinfo{author}{\bibfnamefont{F.~H.} \bibnamefont{Stillinger}},
  \bibinfo{journal}{Science} \textbf{\bibinfo{volume}{209}},
  \bibinfo{pages}{451 } (\bibinfo{year}{1980}).

\bibitem[{\citenamefont{Widom}(1967)}]{Widom:67}
\bibinfo{author}{\bibfnamefont{B.}~\bibnamefont{Widom}},
  \bibinfo{journal}{Science} \textbf{\bibinfo{volume}{157}},
  \bibinfo{pages}{375} (\bibinfo{year}{1967}).

\bibitem[{\citenamefont{Hummer et~al.}(1996)\citenamefont{Hummer, Garde,
  Garc\'{i}a, Pohorille, and Pratt}}]{HummerG:Anitm}
\bibinfo{author}{\bibfnamefont{G.}~\bibnamefont{Hummer}},
  \bibinfo{author}{\bibfnamefont{S.}~\bibnamefont{Garde}},
  \bibinfo{author}{\bibfnamefont{A.~E.} \bibnamefont{Garc\'{i}a}},
  \bibinfo{author}{\bibfnamefont{A.}~\bibnamefont{Pohorille}},
  \bibnamefont{and} \bibinfo{author}{\bibfnamefont{L.~R.} \bibnamefont{Pratt}},
  \bibinfo{journal}{Proc. Natl. Acad. Sci. USA} \textbf{\bibinfo{volume}{93}},
  \bibinfo{pages}{8951 } (\bibinfo{year}{1996}).

\bibitem[{\citenamefont{Ashbaugh and Pratt}(2006)}]{Ashbaugh:2006}
\bibinfo{author}{\bibfnamefont{H.~S.} \bibnamefont{Ashbaugh}} \bibnamefont{and}
  \bibinfo{author}{\bibfnamefont{L.~R.} \bibnamefont{Pratt}},
  \bibinfo{journal}{Rev.\ Mod. \ Phys.} \textbf{\bibinfo{volume}{78}},
  \bibinfo{pages}{159} (\bibinfo{year}{2006}).

\bibitem[{\citenamefont{Hummer et~al.}(2001)\citenamefont{Hummer, Rasaiah, and
  Noworyta}}]{HummerG:Watcth}
\bibinfo{author}{\bibfnamefont{G.}~\bibnamefont{Hummer}},
  \bibinfo{author}{\bibfnamefont{J.~C.} \bibnamefont{Rasaiah}},
  \bibnamefont{and} \bibinfo{author}{\bibfnamefont{J.~P.}
  \bibnamefont{Noworyta}}, \bibinfo{journal}{Nature}
  \textbf{\bibinfo{volume}{414}}, \bibinfo{pages}{188 } (\bibinfo{year}{2001}).

\bibitem[{\citenamefont{Rowlinson and Swinton}(1982)}]{Rowlinson}
\bibinfo{author}{\bibfnamefont{J.~S.} \bibnamefont{Rowlinson}}
  \bibnamefont{and} \bibinfo{author}{\bibfnamefont{F.~L.}
  \bibnamefont{Swinton}}, \emph{\bibinfo{title}{Liquids and Liquid Mixtures}}
  (\bibinfo{publisher}{Butterworths}, \bibinfo{address}{NY},
  \bibinfo{year}{1982}).

\bibitem[{\citenamefont{Asthagiri et~al.}(2006)\citenamefont{Asthagiri,
  Ashbaugh, Piryatinski, Paulaitis, and Pratt}}]{CF4}
\bibinfo{author}{\bibfnamefont{D.}~\bibnamefont{Asthagiri}},
  \bibinfo{author}{\bibfnamefont{H.~S.} \bibnamefont{Ashbaugh}},
  \bibinfo{author}{\bibfnamefont{A.}~\bibnamefont{Piryatinski}},
  \bibinfo{author}{\bibfnamefont{M.~E.} \bibnamefont{Paulaitis}},
  \bibnamefont{and} \bibinfo{author}{\bibfnamefont{L.~R.} \bibnamefont{Pratt}},
  \bibinfo{type}{Tech. Rep.}, \bibinfo{institution}{Los Alamos Natl. Lab.
  {LA}-{UR}-06-3812} (\bibinfo{year}{2006}).

\bibitem[{\citenamefont{Asthagiri et~al.}(2003)\citenamefont{Asthagiri, Pratt,
  and Kress}}]{Asthagiri:2003}
\bibinfo{author}{\bibfnamefont{D.}~\bibnamefont{Asthagiri}},
  \bibinfo{author}{\bibfnamefont{L.~R.} \bibnamefont{Pratt}}, \bibnamefont{and}
  \bibinfo{author}{\bibfnamefont{J.~D.} \bibnamefont{Kress}},
  \bibinfo{journal}{Phys. \ Rev.} \textbf{\bibinfo{volume}{68}},
  \bibinfo{pages}{041505} (\bibinfo{year}{2003}).

\bibitem[{\citenamefont{Paliwal et~al.}(2006)\citenamefont{Paliwal, Asthagiri,
  Pratt, Ashbaugh, and Paulaitis}}]{PaliwalA.:Anamp}
\bibinfo{author}{\bibfnamefont{A.}~\bibnamefont{Paliwal}},
  \bibinfo{author}{\bibfnamefont{D.}~\bibnamefont{Asthagiri}},
  \bibinfo{author}{\bibfnamefont{L.~R.} \bibnamefont{Pratt}},
  \bibinfo{author}{\bibfnamefont{H.~S.} \bibnamefont{Ashbaugh}},
  \bibnamefont{and} \bibinfo{author}{\bibfnamefont{M.~E.}
  \bibnamefont{Paulaitis}}, \bibinfo{journal}{J. Chem. Phys.}
  \textbf{\bibinfo{volume}{124}} (\bibinfo{year}{2006}).

\bibitem[{\citenamefont{Paulaitis and Pratt}(2002)}]{Paulaitis:2002}
\bibinfo{author}{\bibfnamefont{M.~E.} \bibnamefont{Paulaitis}}
  \bibnamefont{and} \bibinfo{author}{\bibfnamefont{L.~R.} \bibnamefont{Pratt}},
  \bibinfo{journal}{Adv. \ Prot.\ Chem.} \textbf{\bibinfo{volume}{62}},
  \bibinfo{pages}{283} (\bibinfo{year}{2002}).

\bibitem[{\citenamefont{Beck et~al.}(2006)\citenamefont{Beck, Paulaitis, and
  Pratt}}]{Beck:2006}
\bibinfo{author}{\bibfnamefont{T.~L.} \bibnamefont{Beck}},
  \bibinfo{author}{\bibfnamefont{M.~E.} \bibnamefont{Paulaitis}},
  \bibnamefont{and} \bibinfo{author}{\bibfnamefont{L.~R.} \bibnamefont{Pratt}},
  \emph{\bibinfo{title}{The potential distribution theorem and models of
  molecular solutions}} (\bibinfo{publisher}{Cambridge University Press},
  \bibinfo{year}{2006}).

\bibitem[{\citenamefont{Hummer et~al.}(1995)\citenamefont{Hummer, Pratt, and
  Garc\'{i}a}}]{Hummer:1995}
\bibinfo{author}{\bibfnamefont{G.}~\bibnamefont{Hummer}},
  \bibinfo{author}{\bibfnamefont{L.~R.} \bibnamefont{Pratt}}, \bibnamefont{and}
  \bibinfo{author}{\bibfnamefont{A.~E.} \bibnamefont{Garc\'{i}a}},
  \bibinfo{journal}{J. \ Phys. \ Chem.} \textbf{\bibinfo{volume}{99}},
  \bibinfo{pages}{14188} (\bibinfo{year}{1995}).

\bibitem[{\citenamefont{Hummer et~al.}(1997)\citenamefont{Hummer, Pratt, and
  Garc\'{i}a}}]{Hummer:1997}
\bibinfo{author}{\bibfnamefont{G.}~\bibnamefont{Hummer}},
  \bibinfo{author}{\bibfnamefont{L.~R.} \bibnamefont{Pratt}}, \bibnamefont{and}
  \bibinfo{author}{\bibfnamefont{A.~E.} \bibnamefont{Garc\'{i}a}},
  \bibinfo{journal}{J. \ Am. \ Chem. \ Soc.} \textbf{\bibinfo{volume}{119}},
  \bibinfo{pages}{8523} (\bibinfo{year}{1997}).

\bibitem[{\citenamefont{Pratt and Asthagiri}(2006)}]{CPMS}
\bibinfo{author}{\bibfnamefont{L.~R.} \bibnamefont{Pratt}} \bibnamefont{and}
  \bibinfo{author}{\bibfnamefont{D.}~\bibnamefont{Asthagiri}}, in
  \emph{\bibinfo{booktitle}{Free Energy Calculations. Theory and Applications
  in Chemistry and Biology}}, edited by
  \bibinfo{editor}{\bibfnamefont{C.}~\bibnamefont{Chipot}} \bibnamefont{and}
  \bibinfo{editor}{\bibfnamefont{A.}~\bibnamefont{Pohorille}}
  (\bibinfo{publisher}{Springer}, \bibinfo{address}{Berlin},
  \bibinfo{year}{2006}), chap. \bibinfo{chapter}{9. Potential distribution
  methods and free energy models of molecular solutions}.

\bibitem[{\citenamefont{Hansen and McDonald}(2006)}]{Hansen-McDonald}
\bibinfo{author}{\bibfnamefont{J.-P.} \bibnamefont{Hansen}} \bibnamefont{and}
  \bibinfo{author}{\bibfnamefont{I.~R.} \bibnamefont{McDonald}},
  \emph{\bibinfo{title}{Theory of simple liquids}}
  (\bibinfo{publisher}{Elsevier}, \bibinfo{year}{2006}).

\end{thebibliography}

\end{document}